\documentclass[12pt,preprint]{aastex}

\def\gs{\mathrel{\raise0.35ex\hbox{$\scriptstyle >$}\kern-0.6em
\lower0.40ex\hbox{{$\scriptstyle \sim$}}}}
\def\ls{\mathrel{\raise0.35ex\hbox{$\scriptstyle <$}\kern-0.6em
\lower0.40ex\hbox{{$\scriptstyle \sim$}}}}

\shorttitle{A cosmic ray-dominated ISM in ULIRGs}

\begin{document}

\title{A cosmic-ray dominated ISM in Ultra Luminous Infrared Galaxies: new initial conditions
for star formation}

\author{Padelis P. Papadopoulos\altaffilmark{1}}
\altaffiltext{1} {Argelander-Institut f\"ur Astronomie,  Auf dem H\"ugel 71, 
  D-53121 Bonn, Germany}

\begin{abstract}

  The  high-density  star formation  typical  of the  merger/starburst
 events  that  power  the  large  IR luminosities  of  Ultra  Luminous
 Infrared Galaxies (ULIRGs) ($\rm L_{IR}$(8--1000\,$\mu$m)$\ga
 $$10^{12}$\,L$_{\odot}$)   throughout   the   Universe   results   to
 extraordinarily  high cosmic  ray  (CR) energy  densities of  U$_{\rm
 CR}$$\sim $few$\times  $($10^3$--$10^4$)\,U$_{\rm CR,Gal}$ permeating
 their interstellar  medium (ISM), a  direct consequence of  the large
 supernovae remnants (SNRs) number  densities in such systems.  Unlike
 far-UV  photons emanating  from  their numerous  star forming  sites,
 these large  CR energy densities  in ULIRGs will  volumetrically heat
 and    raise    the    ionization    fraction    of    dense    ($\rm
 n$$>$10$^4$\,cm$^{-3}$)   UV-shielded  gas  cores   throughout  their
 compact star-forming  volumes.  Such conditions can turn  most of the
 large  molecular gas  masses found  in  such systems  and their  high
 redshift  counterparts  ($\sim $10$^9$--10$^{10}$\,M$_{\odot}$)  into
 giant   CR-dominated  Regions  (CRDRs)   rather  than   ensembles  of
 Photon-dominated  Regions (PDRs) which  dominate in  less IR-luminous
 systems where star formation and molecular gas distributions are much
 more extended.  The molecular gas  in CRDRs will have a {\it minimum}
 temperature  of  $\rm  T_{kin}$$\sim  $(80--160)\,K,  and  very  high
 ionization fractions  of x(e)$>$10$^{-6}$ throughout  its UV-shielded
 dense cores, which in turn  will {\it fundamentally alter the initial
 conditions for star formation  in such systems.}  Observational tests
 of CRDRs can be provided by  high-J CO and $^{13}$CO lines or multi-J
 transitions  of  any heavy  rotor  molecules  (e.g.   HCN) and  their
 isotopologues.  Chemical signatures of very high ionization fractions
 in dense UV-shielded  gas such as low $\rm  [DCO^+]/[HCO^+]$ and high
 $\rm [HCO^+]/[CO]$ abundance ratios would  be good probes of CRDRs in
 extreme starbursts.   These tests, along with  direct measurements of
 the high CO  line brightness temperatures expected over  the areas of
 compact  dense gas disks  found in  ULIRGs will  soon be  feasible as
 sub-arcsecond interferometric imaging capabilities and sensitivity at
 mm/submm wavelengths improve in the era of~ALMA.

\end{abstract}

\keywords{galaxies: star formation -- galaxies: starburst -- galaxies: IR -- ISM: molecules --
ISM: dust -- ISM: cosmic rays}

\section{Introduction}

Cosmic  rays (CRs) have  been established  for some  time as  the main
regulators of the temperature, ionization, and chemical state of dense
gas cores lying deep inside the (far-UV)-shielded regions of molecular
clouds (e.g.   Goldsmith \& Langer 1978; Goldsmith  2001; Lequeux 2004
and  references  therein).   The  association  of CRs  with  O,B  star
clusters and supernovae remnants (SNRs) where they are accelerated (in
massive-star winds  and SN shocks) has been  recently demonstrated for
the Galaxy  (Binns et al.  2008),  and even shown  for individual SNRs
(Acciari et  al.  2009a)  and star-forming (SF)  regions (Abdo  et al.
2010),  while   the  synchrotron  emission   of  CR  electrons   is  a
well-established marker  of star-forming regions (e.g.   Condon et al.
1990, 1991).  The recent detections  of $\gamma $-rays (the product of
inelastic  collisions between  CR-protons and  hydrogen nuclei  in the
ISM) from  the starburst nuclei  of M\,82 and NGC\,253  solidified the
connection of CR energy density to SNRs and star formation activity in
galaxies (Acciari et al.  2009b; Acero et al.  2009).  CRs rather than
far-UV photons have even been  advocated as the {\it dominant} heating
mechanism  of molecular  gas in  galaxies irrespective  of  their star
formation  activity (Suchkov,  Allen, \&  Heckman 1993  and references
therein).  These early proposals  however faced two distinct problems:
a)  ensembles of  standard Photon  Dominated Regions  (PDRs) accounted
well for the global molecular  and atomic line emission from quiescent
spirals  such as  the Milky  Way (Fixsen  et al.   1999;  Mochizuki \&
Nakagawa 2000) as well as  starbursts (e.g.  Wolfire et al.  1990; Mao
et al.  2000),  and b) a tight correlation  between CO line brightness
and  non-thermal radio continuum  used as  evidence for  CR-heating of
molecular gas  can be also  attributed to the  well-known far-IR/radio
correlation.   The latter  is established  by star  formation powering
both the far-IR  dust continuum and the non-thermal  radio emission in
galaxies, with far-UV photons from SF sites heating the dust (and thus
the  gas via photoelectric  heating) and  reprocessed into  the far-IR
continuum  (e.g.   Condon  \&  Yin  1990;  Condon  1992).   The  tight
far-IR/radio correlation  and the close  association of O,B  stars and
SNRs (the  sites of both far-UV  photons and CR  acceleration) down to
individual CO-luminous  Giant Molecular Clouds (GMCs)  can then easily
account for the  observed (CO intensity)/(radio continuum)~correlation
without CR-heating  of CO-bright clouds  as an underlying  cause (i.e.
the far-IR/radio becomes a CO/radio  correlation as CO luminosity is a
good proxy for far-IR luminosity).

It   was  the   recent  capability   for  sensitive   observations  of
high-excitation high-J transitions of  CO and $^{13}$CO that ``broke''
the  aforementioned  degeneracies  and definitively  demonstrated  the
presence of  CR-heated molecular gas  in the starburst nucleus  of the
otherwise  quiescent spiral  galaxy NGC\,253  (Bradford et  al.  2003;
Hailey-Dunsheath et al.  2008), while strong evidence suggests this is
also the case for the dense molecular cloud in Sgr\,B2 at the Galactic
Center (Yusef-Zadeh  et al. 2007).  Nevertheless these  amount to only
$\sim $0.1--1\%  of the total  molecular gas found in  typical spirals
and thus  do not  change the  paradigm of far-UV  photons as  the main
heating agent  of molecular gas  in IR-luminous galaxies.  It  must be
noted that irrespective whether CRs  heat most of the molecular gas in
galaxies or not, they remain the ultimate regulator of its temperature
and  ionization   fraction  minima,  both  reached   in  its  densest,
UV-shielded, phase deep inside molecular~clouds.

\section{The CR energy density in ULIRGs: a CR-dominated ISM}

It  is the star  formation rate  (SFR) {\it  density} rather  than the
total  SFR  that  determines   the  CR  energy  density  $\rm  U_{CR}$
(eV\,cm$^{-3}$) in  the ambient ISM  of a galaxy.   Indicatively, for
the  100\,pc  starburst  nucleus  of  NGC\,253  the  latter  is  $\sim
$$2\times  10^3$   times  higher   than  the  Galactic   value  ($\sim
$0.5\,eV\,cm$^{-3}$) even though their global SFRs are similar (Accero
et al.  2009).  In the central 500\,pc of the nearby starbust M\,82 it
is  $\rm U_{CR}$$\sim  $500$\times $$\rm  U_{CR,Gal}$ (Suchkov  et al.
1993; Acciari  et al.   2009b) while its  globally averaged  CR energy
density remains  similar to that of  the Galaxy.  Such  high values of
$\rm U_{CR}$  can be  easily attained and  surprassed over  the entire
volume  of   the  substantial  reservoirs  of   molecular  gas  ($\sim
$(10$^9$--10$^{10}$)\,M$_{\odot}$) in Ultra Luminous Infrared Galaxies
(hereafter ULIRGs)  and their high redshift  counterparts.  Indeed the
large   IR    luminosities   of   these    extreme   starbusts   ($\rm
L_{IR}$(8--1000\,$\mu  $m)$\ga  $10$^{12}$\,L$_{\odot}$) emanate  from
very small volumes with typical IR brightnesses of

\begin{equation}
\rm  \sigma _{IR}(ULIRGs) = \left(10^{12}-10^{13}\right)\, \frac{L_{\odot}}{kpc^2},
\end{equation}

\noindent
and   with  most   such  systems   strongly  clustering   around  $\rm
\sigma_{IR}$$\sim   $10$^{13}$\,L$_{\odot  }$\,kpc$^{-2}$,  (Thompson,
Quataert,  \& Murray  2005).  This  latter value  could  be indicating
radiation-pressure regulated maximal starbusts (Thompson, Quataert, \&
Murray 2005; Thompson  2009), where an Eddington limit  from O, B star
clusters sets a  maximum gas accretion rate onto  SF sites deep inside
molecular clouds via photon pressure on its concomitant dust (see also
Scoville 2004 for an  earlier and simple exposition).  Interestingly a
similar threshold value  for $\rm \sigma _{IR}$ can  be also recovered
with  CRs instead of  photons setting  the Eddington  limit (Socrates,
Davis, \&  Ramirez-Ruiz 2008) as CRs  are much more  highly coupled to
the ISM  than photons.  Either  way the high $\rm  \sigma_{IR}$ values
typical in  ULIRGs seem  to be the  result of {\it  extreme starbursts
occuring in very compact  regions,} while for less IR-luminous systems
($\rm L_{IR}$$\la $10$^{11}$\,L$_{\odot}$),  which typically have more
extended        star        formation:     $\rm        \sigma_{IR}$=
(10$^{10}$--10$^{11}$)\,L$_{\odot}$\,kpc$^{-2}$  (Lehnert  \&  Heckman
1996).   For nearby  ULIRGs this  compactness of  their  molecular gas
reservoirs  (and   thus  of  their  star-forming   volumes)  has  been
demonstrated using mm and  (recently) submm interferometric imaging of
their CO line and dust continuum, that found gaseous disks with D$\sim
$(100--300)\,pc  (e.g.  Downes et  al.  1998;  Sakamoto et  al.  2008;
Matsushita et al.  2009).   Moreover, even when such systems initially
seem   as  having   larger   dimensions,  higher-resolution   mm/submm
interferometry  frequently reveals  them  as mergers  of such  compact
nuclei,  each containing the  molecular gas  of a  gas-rich progenitor
with  M(H$_2$)$\sim $(10$^9$--10$^{10}$)\,M$_{\odot  }$ (Evans  et al.
2002; Sakamoto et al.  2008).

 Recently  $\rm \sigma  _{IR} $  values similar  to those  typical for
local ULIRGs have been  found also for a submillimeter-selected galaxy
(SMG)  at z$\sim $2.3  where a  unique combination  of high-resolution
submm imaging and a strong magnification by gravitational lensing made
possible the resolution of the star-forming area of this distant ULIRG
at linear scales of $\sim  $100\,pc (Swinbank et al.  2010).  Finally,
while  the level  of the  contribution of  an Active  Galactic Nucleus
(AGN) to  the tremendeous  $\rm \sigma _{IR}$  values of  such compact
regions remains a  matter of debate (e.g.  Downes  \& Eckart 2007), it
can  be   safely  assumed   that  they  are   good  order-of-magnitude
``calorimeters''  of  dust-obscured   SFR  (e.g.   in  the  archetypal
QSO/starburst  system Mrk\,231:  $\rm L_{IR}$=2/3(starburst)+1/3(AGN),
Downes  \& Solomon  1998).  This  is  further supported  by the  large
number of radio SNRs found recently within the gaseous disks of ULIRGs
using VLBI imaging (Sakamoto et al.  al.  2008 and references therein;
P\'erez-Torres et al. 2009),  while similar AGN contributions to total
IR  luminosities  are recovered  also  for  dusty  starbursts at  high
redshifts (Pope et al.  2008; Murphy et al. 2009).

The IR luminosity surface density  and the SN rate surface density can
then be related using the well-established IR/radio correlation

\begin{equation}
\rm \frac{[\nu (GHz) L_{\nu}]}{L_{IR}} = \mu (\nu)\, 10^{-6}
\end{equation}

\noindent
(Yun et al. 2001 and  references therein). This relates the power $\rm
\nu L_{\nu }$ of the  non-thermal radio continuum to the IR luminosity
(and  holds over  five  orders  of magnitude  with  a dispersion  $\la
80\%$), with the star formation  rate consider as the ultimate scaling
factor for both IR and non-thermal radio luminosities (see Thompson et
al.  2006 for the latest theoretical background).  To obtain a working
value for  $\mu(\nu)$ we follow Yun  et al.  2001 and  use their value
for the  far-IR/radio correlation  at $\nu $=1.4\,GHz:  $\rm \left(\nu
L_{\nu}\right)/L_{FIR}$=$\rm 2.66\times  10^{-4}\times 10^{-q}$, where
q=2.35  (as  defined  by  Helou  et  al.   1985).   For  $\langle  \rm
L_{IR}$(8--1000\,$\mu $m)/$\rm L_{FIR}$(42--120\,$\mu $m)$\rangle$=1.3
(Yun et  al.  2001),  the coefficient in  Equation 2 becomes  $\rm \mu
(\nu)$=1.54,  and  is  adopted   for  this  study  (with  an  inherent
uncertainty  of  a   factor  of  $\sim  $2  due   to  a  varying  $\rm
L_{FIR}/L_{IR}$  in  IR-luminous galaxies).   On  the  other hand  the
non-thermal radio power depends on the SN rate $\rm f_{SN}$ as

\begin{equation}
\rm \frac{L_{\nu }}{10^{22}\,W\,Hz^{-1}}= 13\,\left[\nu(GHz)\right]^{-\alpha} f_{SN}(yr^{-1}),
\end{equation}

\noindent
(Condon  1992).  Combining  Equations   2  and  3  and  assuming  full
concomitance  of  IR and  non-thermal  radio  emission  over the  star
forming regions yields the SN rate surface density

\begin{equation}
\rm \frac{\dot\Sigma _{SN}}{yr^{-1}\,kpc^{-2}}=2.95 \mu(\nu)
\left(\frac{\sigma_{IR}}{10^{12}\,L_{\odot}\,kpc^{-2}}\right) [\nu(GHz)]^{\alpha -1}.
\end{equation}

 Following Suckhov  et al.  (1993)  the CR energy densities  in ULIRGs
scale   with   respect   to   that   of the Galaxy as

\begin{equation}
\rm \frac{U_{CR}}{U_{CR,Gal}}\sim \frac{\dot\Sigma _{SN}}{\dot\Sigma _{SN,Gal}}\times
\left(\frac{V_{diff}}{V_{wind}}\right), 
\end{equation}

\noindent
where               $\rm              \dot\Sigma_{SN,Gal}$=3.85$\times
$10$^{-5}$\,yr$^{-1}$\,kpc$^{-2}$ (McKee  \& Williams 1997),  and $\rm
V_{diff}$ is the diffusion velocity at which CRs escape from quiescent
disks like  the Milky Way  while $\rm V_{wind}$  is the velocity  of a
SF-induced wind  at which CRs  are advected out  of the SF  regions in
starbursts.  Typically $\rm V_{diff}$$\sim $10\,km\,s$^{-1}$ while the
maximum    velocity   of    SF-driven   galactic    winds    is   $\rm
V_{wind,max}$$\sim $(2--3)$\times $10$^3$\,km\,s$^{-1}$ (Suchkov et a.
1993;  Veilleux  et  al.   2005).  Such  starburst-induced  high  wind
velocities  are  deduced for  M\,82  (Seaquist  et  al.  1985)  though
usually   $\rm    V_{wind}$$\sim   $(500--1000)\,km\,s$^{-1}$.    From
Equations       4      and       5,      after       setting      $\rm
V_{wind,max}$=3000\,km\,s$^{-1}$, and $\mu (\nu)$$\sim $ 1.54, $\alpha
$=0.8  (for $\nu$=1.4\,GHz),  Equation~1 yields  CR energy  density in
compact starbursts (CSB) of

\begin{equation}
\rm \frac{U_{CR,CSB}}{U_{CR,Gal}}\sim 4\times(10^2-10^4).
\end{equation}

\noindent
For    ULIRGs    as   the    template    compact   starbursts:    $\rm
  \sigma_{IR}$=10$^{13}$\,L$_{\odot}$\,kpc$^{-2}$,    yielding    $\rm
  U_{CR,CSB}$$\sim 4\times 10^3\rm U_{CR,Gal}$.  The latter amounts to
  a  tremendeous boost  of CR  energy density,  similar to  that found
  recently for  a single  molecular cloud in  the nucleus  of NGC\,253
  (Acero et al. 2009), and is capable of turning the massive and dense
  molecular  clouds  of ULIRGs  into  CR-dominated Regions  (hereafter
  CRDRs) in terms of the  dominant heating mechanism.  For the extreme
  starbursts in ULIRGs and their high redshift counterparts these high
  $\rm \sigma_{IR}$  and correspondingly high $\rm  U_{CR}$ values can
  involve  $\sim $(10$^9$--10$^{10}$)\,M$_{\odot  }$ of  molecular gas
  mass fueling galaxy-wide star-forming episodes.

\subsection{CRDRs in ULIRGs: a new  temperature minimum for  UV-shielded gas}

For the UV-shielded, and mostly  subsonic, dense gas cores deep inside
molecular clouds the heating rate is given by

\begin{equation}
\rm \Gamma _{CR}\sim 1.5\times 10^{-24}\left(\frac{\zeta_{CR}}{10^{-17}\,s^{-1}}\right)
\left(\frac{n(H_2)}{10^4\,cm^{-3}}\right)\,erg\,cm^{-3}\,s^{-1}, 
\end{equation}

\noindent 
where  $\rm  \zeta_{CR}$(s$^{-1}$)$\rm   \propto  U_{CR}$  is  the  CR
ionization  rate per  H$_2$  molecule. The  gas  cooling via  gas-dust
interaction can be expressed as

\begin{equation}
\rm \Lambda _{g-d}\sim 10^{-25}\left(\frac{n(H_2)}{10^4\,cm^{-3}}\right)^2 T^{1/2} _k
\left(T_k-T_{dust}\right)\,erg\,cm^{-3}\,s^{-1}
\end{equation}

\noindent
(see  Tielens  2005 for  derivation  of  both  expressions). The  most
important line cooling  of this gas phase is  through rotational lines
of CO and a few other molecular species whose rotational ladder energy
levels  reaches  down  to  $\rm \Delta  E_{ul}/k_B$$\sim  $(5--10)\,K.
Following the  detailed study of  Goldsmith (2001) we  parametrize the
molecular line cooling as

\begin{equation}
\rm \Lambda _{line}\sim 6\times 10^{-24}\left[\frac{n(H_2)}{10^4\,cm^{-3}}\right]^{1/2}
\left(\frac{T_{k}}{10\,K}\right)^{\beta} erg\,cm^{-3}\,s^{-1}.
\end{equation}

\noindent
The  density dependence  was extracted  from  a fit  of the  parameter
$\alpha $ in Table 2 of Goldsmith (2001), and reproduces the values of
the   $\rm  \Lambda   _{line}$   to  within   $\la   20\%$  for   $\rm
n(H_2)$=(10$^4$--10$^6$)\,cm$^{-3}$, which spans  the density range of
dense cores within GMCs. For that density range $\beta $$\sim $3.

The  resulting gas  temperature  for  the CR-heated  gas  can then  be
estimated from the equation of thermal balance

\begin{equation}
\rm \Gamma _{CR} = \Lambda_{line}+\Lambda_{g-d}.
\end{equation}

\noindent
Setting  the dust  temperature  to $\rm  T_{dust}$=0\,K  will yield  a
minimum  $\rm  T_{k}$ value  while  also  allowing  a simple  analytic
solution of  Equation 10. Substituting the  expressions from Equations
7, 8 and 9 into the latter yields

\begin{equation}
\rm T^3 _{k,10}+0.526n^{3/2} _4T^{3/2} _{k,10}=0.25n^{1/2}_4\zeta_{-17},
\end{equation}

\noindent
where        $\rm       T_{k,10}$=$\rm        T_k$/(10\,K),       $\rm
n_4$=n(H$_2$)/(10$^4$\,cm$^{-3}$)   and   $\zeta  _{-17}$=$\rm   \zeta
_{CR}/(10^{-17}\,s^{-1})$. An exact solution of the latter then is

\begin{equation}
\rm T_{k,10}=0.630\left[\left(n^{1/2}_4 \zeta_{-17}+0.276676n^3_4\right)^{1/2}-0.526n^{3/2}_4\right]^{2/3}
\end{equation}

  For  $\rm \zeta _{CR,  Gal}$=5$\times $10$^{-17}$\,s$^{-1}$  for the
Galaxy   (e.g.    van   der    Tak   \&   van   Dishoeck   2000)   and
n(H$_2$)=10$^4$\,cm$^{-3}$  the latter  yields $\rm  T_{k}$$\sim $9\,K
which is  typical for UV-shielded  gas immersed in Galactic  CR energy
density (e.g.   Goldsmith 2001), and deduced  by numerous observations
in the  Galaxy (e.g. Pineda \&  Bensch 2007; Bergin  \& Tafalla 2007).
For  $\rm \zeta_{CR,  CSB}$=(1--4)$\times  $10$^3$$\rm \zeta_{CR,Gal}$
expected   for    the   ISM   of   extreme    starbursts,   and   $\rm
n(H_2)$=(10$^4$--10$^{6}$)\,cm$^{-3}$  (typical  densities  for  dense
cores in the Galaxy, e.g.   Bergin \& Tafalla 2007) Equation 12 yields
$\rm T_k$$\sim $(80--240)\,K, as {\it the minimum possible temperature
for the molecular  gas in compact extreme starbursts}  (see Figure 1).
Turbulent gas  heating (e.g.   Pan \& Padoan  2009) that  may ``seep''
down  the molecular  cloud hierarchical  structures to  the  dense gas
cores  (although as  typically  subsonic, they  are  expected to  have
neglible such  heating), any sort  of mechanical heating of  the dense
gas in ULIRGs  (Baan, Loenen, \& Spaans 2010),  or warmer dust because
of IR  light ``leaking''  deep inside molecular  clouds {\it  can only
raise this temperature range}.


 The gas cores  deep inside the CR-heated regions  of molecular clouds
in  the  Galaxy,  and  especially  the  highest  density  ones  ($\sim
$(10$^5$--10$^6$)\,cm$^{-3}$) are  typically dominated by near-thermal
motions  (e.g.  Bergin  \&  Tafalla 2007).   However  the often  large
turbulent linewidths  found in  the dense gas  disks of nearby ULIRGs
(Downes  \& Solomon  1998; Sakamoto  et al.   2008, Matsushita  et al.
2009) could  in principle affect  even dense  gas core  kinematics and
thus their line cooling function.  Following Goldsmith (2001) that any
such  macroscopic motions  are driven  mostly by  self-gravity,  a new
effective line cooling function would be

\begin{equation}
\rm \Lambda ^{(eff)} _{line}= \Lambda _{line}\times \left(\frac{n(H_2)}{10^3\,cm^{-3}}\right)^{1/2},
\end{equation}

\noindent
which rougly quantifies the effect of increased transparency (and thus
cooling power) of molecular  line photons. An optical depth dependance
of  $\rm  \tau  $$\propto   $(dV/dR)$^{-1}$  is  adopted,  where  $\rm
(dV/dR)_{VIR}$=$\rm
\left[n(H_2)/(10^3\,cm^{-3})\right]^{1/2}$\,km\,s$^{-1}$\,pc$^{-1}$ is
the velocity gradient of macroscopic motions for self-gravitating gas.
The equation of thermal balance and its solution then become

\begin{equation}
\rm T^3_{k,10}+0.166n_4 T^{3/2} _{k,10} = 0.0789\zeta_{-17},
\end{equation}

\noindent
and 

\begin{equation}
\rm T_{k,10} = 0.630\left[\left(0.02766n^2_4 + 0.3156\zeta_{-17}\right)^{1/2}-0.1663n_4\right]^{2/3},
\end{equation}

\noindent
(where we set again  $\rm T_{dust}$=0\,K).  For $\rm \zeta _{CR}$$\sim
$(1--4)$\times  $10$^3$$\rm \zeta _{CR,Gal}$  the new  thermal balance
equation  yields $\rm T_k$(min)$\sim  $(55--115)\,K for  the molecular
gas in  such environments  while for the  more extreme CSBs  with $\rm
\zeta   _{CR}$$\sim   $10$^4$$\rm   \zeta   _{CR,Gal}$  it   is   $\rm
T_k$(min)$\sim $(145--160)\,K.  In  all cases the minimum temperatures
in  CRDRs  remain  significantly  higher the  minimum  $\rm  T_k$$\sim
$(8--10)\,K which  is attained  in the dark  UV-shielded cores  in the
Galaxy (see Figure  1) where the initial conditions  of star formation
are set (Bergin \& Tafalla 2007).

It must  be noted that such  high CR energy densities  may also affect
molecular gas  chemistry, and thus  the abundance of coolants  such as
CO.  Hence  better temperature estimates of the  UV-shielded dense gas
cores in  CRDRs can only  be provided by self-consistent  solutions of
their thermal {\it and}  chemical states. Moreover activation of other
cooling lines  such as OI  at 63$\mu $m ($\rm  \Delta E_{ul}/k_B$$\sim
$228\,K) when  temperatures rise  significantly above 100\,K  in dense
gas  ($\rm  n(H_2)\ga  10^5$\,cm$^{-3}$)  can  cap the  rise  of  $\rm
T_k$(min)=F($\rm  \zeta  _{CR}$)  in   CRDRs  to  $\sim  150$\,K  (Thi
2010). One the other hand gas temperatures can be even higher than the
simple estimates provided by Equation 10 because of residual turbulent
motion dissipation  (and thus heating)  remains possible in  the dense
gas  reservoirs of ULIRGs,  as they  may resemble  the tidally-stirred
dense  gas in  the  Galactic Center  where  turbulent heating  remains
important   even  at  high   densities  (e.g.    Stark  et   al  1989;
Rodriguez-Fernandez et al. 2001; G\"usten \& Philipp~2004).

\subsection{The ionization fraction of dense gas in ULIRGs}

The large  cosmic ray energy densities,  besides significantly raising
the  minimum possible  temperature  of molecular  gas  in the  compact
starbursts powering ULIRGs and  their high redshift counterparts, they
will  also   dramatically  raise  the   minimum  ionization  fraction.
Following the  treatment by McKee (1989),  in UV-shielded environments
with  negligible photoionization  and CRs  as  the sole  cause of  ISM
ionization,     the     ionization     fraction    $\rm     x(e)$=$\rm
n_e/n(H)=n_e/2n(H_2)$ is given by

\begin{equation}
\rm x(e) = 2\times 10^{-7} r^{-1} _{gd} \left(\frac{n_{ch}}{2n(H_2)}\right)^{1/2} 
\left[\left(1+\frac{n_{ch}}{8n(H_2)}\right)^{1/2} + \left(\frac{n_{ch}}{8n(H_2)}
\right)^{1/2}\right],
\end{equation}

\noindent
where   $\rm   n_{ch}$$\sim   $500\,$\rm   \left(r^2   _{gd}\,   \zeta
_{-17}\right)\, cm^{-3}$ is a characteristic density encapsulating the
effect  of  cosmic rays  and  ambient  metallicity  on the  ionization
balance  ($\rm   r_{gd}$  is   the  normalized  gas/dust   ratio  $\rm
r_{gd}$=[(G/D)/100] with  G/D(gas-to-dust mass)=100 assumed  for Solar
metallicities; e.g. Knapp \& Kerr 1974; Aannestad \& Purcell 1973).

For            the             Galaxy            where            $\rm
\zeta_{CR,Gal}$=5$\times$10$^{-17}$\,s$^{-1}$      it      is     $\rm
n_{ch}$=2.5$\times  $10$^3$\,cm$^{-3}$, and for  a typical  dense core
density  of $\rm n(H_2)$=10$^{5}$\,cm$^{-3}$  Equation 16  yields $\rm
x(e)$$\sim $2.4$\times  $10$^{-8}$, consistent with  the typical range
in the Galaxy: $5\times 10^{-9} \la x(e) \la 1.5 \times 10^{-7}$ (e.g.
Langer  1985).    For  the  much  larger  CR   ionization  rates  $\rm
\zeta_{CR,CSB}$=10$^3$$\times $$\rm  \zeta _{CR,Gal}$ expected  in the
ISM of compact  starbursts: $\rm n_{ch}$=2.5$\times $10$^6$\,cm$^{-3}$
and $\rm  x(e)$$\sim $4$\times $10$^{-6}$.   The latter is one  to two
orders of magnitude  larger than the typical range  and $\sim $4 times
higher  than the highest  value measured  for dense  UV-shielded cores
anywhere  in  the Galaxy  (Caselli  et  al.   1998).  In  the  classic
photoinization-regulated  star formation  scenario  (McKee 1989)  such
high ionization fractions  will be capable of turning  even very dense
gas  in   the  UV-shielded/CR-ionized  regions   of  molecular  clouds
subcritical    (i.e.    $\rm    M_{core}/M_{\Phi}$$<$1    where   $\rm
M_{\Phi}$=$\rm 0.12\Phi/G^{1/2}$,  and $\Phi  $ is the  magnetic field
flux threading  the molecular core, Mouschovias \&  Spitzer 1976), and
thus  halt  their  gravitational  collapse  until a  now  much  slower
ambiplolar diffusion allows it.

\section{Important consequences and some key observational tests}

 The CR-permeated molecular gas  in compact extreme starbursts is more
  than  the mere  sum  of individual  star-forming  regions and  their
  localized dense PDRs.  The latter  would leave most of the dense gas
  settle to  a cold state  since for the  high gas densities  found in
  ULIRGs,  far-UV field  intensities  will be  reduced  by factors  of
  $\sim$10$^4$ over distances of $\la $0.1\,pc.  On the other hand, by
  dramatically  altering the  thermal  and ionization  state of  dense
  UV-shielded  gas cores  inside molecular  clouds {\it  the  large CR
  energy  densities  in  extreme  starbusts  significantly  alter  the
  initial conditions  for star formation in such  systems.}  Indeed it
  is in the UV-shielded dense gas cores where these initial conditions
  are  set,  and  the  large  temperatures expected  for  these  cores
  throughout  CRDRs invalidates  the  main arguments  about an  almost
  constant characteristic mass of young stars in most ISM environments
  including  starbursts  (Elmegreen,  Klessen,  \& Wilson  2008).  The
  latter study  did not consider the  effects of CRs, and  as a result
  found  that UV-shielded  dense  gas remains  cold ($\rm  T_{k}$$\sim
  $10\,K) even in starbursts.

   The much larger  ionization fractions that can now  be reached deep
  inside molecular clouds can in principle: a) keep the magnetic field
  lines  strongly  ``threaded''  onto  molecular gas  at  much  higher
  densities, and b)  as a result render much of  its mass incapable of
  star-formation, at  least in the  simple photoinization-regulated SF
  scenario.   These effects stem  from the  now much  longer ambipolar
  diffusion timescale

\begin{equation}
\rm \tau _{AD}=3.2\times 10^7 r_{gd} \left(\frac{n_{ch}}{2n(H_2)}\right)^{1/2}
\left[\left(1+\frac{n_{ch}}{8n(H_2)}\right)^{1/2}+\left(\frac{n_{ch}}{8n(H_2)}\right)^{1/2}\right] yrs,
\end{equation}

\noindent
 needed for a CR-ionized dense  core with density $\rm n(H_2)$ to lose
magnetic flux and  collapse (McKee 1989). For the  CR ionization rates
of $\rm \zeta _{CR}$=few$\times $(10$^3$--10$^4$)$\rm \zeta _{CR,Gal}$
expected    in   CRDRs   it    would   be    $\rm   n_{ch}$=2.5$\times
$10$^{6-7}$\,cm$^{-3}$,  thus  for a  typical  dense  core density  of
n(H$_2$)=10$^5$\,cm$^{-3}$,    $\rm    \tau   _{AD}$$\sim    $3$\times
$(10$^8$--10$^{10}$)\,yrs (see Figure 2).  In photoinization-regulated
star  formation this is  a {\it  lower} limit  on the  gas consumption
timescale by  the latter process, and  in CRDRs it  is clearly already
much  longer (by  up  to two  orders  of magnitude)  than the  typical
consumption timescale of molecular gas reservoirs of LIRGs (Figure 2).
For  the more  vigorously  star-forming ULIRGs,  and considering  only
their dense HCN-bright  molecular gas phase as the  true SF fuel (e.g.
Gao         \&         Solomon         2004):        $\rm         \tau
_{cons}$=M(n$>$10$^{4}$\,cm$^{-3}$)/SFR$\sim  $10$^7$\,yrs,  and  $\rm
\tau _{AD}/\tau _{cons}$(n$>$10$^4$\,cm$^{-3}$)$\sim
$$10^2$--$10^3$.

  This      certainly     disfavors     a      simple     quasi-static
photoionization-regulated SF scenario of B-field lines slowly slipping
from stationary dense gas cores  which then proceed to star formation.
Such  a  simple picture  is  expected to  be  modified  anyway by  the
presence  of  MHD  turbulence   which  can  accelarate  the  ambipolar
diffusion  process (Heitsch  et  al.  2004),  especially  in the  very
turbulent  molecular gas  of ULIRGs.   For such  systems  the magnetic
fields can be strong,  in possible equipartition with highly turbulent
gas  (Thompson  et al.   2006),  and  thus  dynamically important  and
strongly co-evolving with the molecular  gas.  It may well be that the
ability of CRDRs  to maintain high ionization fractions  in very dense
UV-shielded molecular  gas, and thus  retain a strong coupling  of the
magnetic  fields  onto   the  bulk  of  its  mass   in  ULIRGs  (where
$\langle$n(H$_2$)$\rangle$$>$10$^{4}$\,cm$^{-3}$),  is  what allows  a
quick  establishment  of  equipartition  between magnetic  fields  and
turbulent   gas  motions   in  their   self-gravitating   disks.   MHD
simulations  for gas  clouds immersed  in the  very intense  CR energy
density backrounds expected  in CRDRs will be key  in addressing these
issues,  and  investigate  whether  turbulence  can  still  accelerate
ambipolar  diffusion  in  molecular  clouds of  such  high  ionization
fractions and lead  back to $\rm \tau _{AD}$$<$$\rm  \tau _{cons}$ for
the dense gas in ULIRGs.

\subsection{A new set of initial conditions for star formation in ULIRGs}
 
In the  high-extinction ISM of compact extreme  starbursts, CRDRs make
the  dramatic  change of  star-formation  initial  conditions {\it  an
imperative  of  high-density   star  formation.}   The  latter  occurs
irrespective whether the bulk of  the molecular gas in CRDRs of ULIRGs
is heated by  CRs or not as the latter will  now raise the temperature
minimum  of pre-stellar  UV-shielded  dense gas  cores throught  their
star-forming volumes by a large  factor.  The consequences of this new
imperative for star-formation in extreme starburst environments remain
invariant irrespective whether  gravitational instability in turbulent
molecular  gas  (e.g Klessen  2004;  Jappsen  et  al.  2005;  Bonnell,
Clarke, \& Bate  2006) or ambipolar diffusion of  magnetic field lines
from dissipated  dense cores followed by  their gravitational collapse
(e.g.  Mouschovias  \& Spitzer 1976, McKee 1989)  drive star formation
in galaxies. In both schemes  the dense UV-shielded cores of molecular
clouds is  where the initial  conditions of star formation  are trully
set  (see Elmegreen  2007;  Ballesteros-Paredes \&  Hartmann 2007  for
recent excellent reviews).  It must  also be noted that this gas phase
is different from the PDR-dominated gas around star-forming sites that
dominates the global molecular line and dust continuum spectral energy
distributions observed for ULIRGs  and often (erroneously) used to set
the star formation initial  conditions in starbursts (e.g.  Klessen et
al.  2007).

  The effects  on the  characteristic mass scale  of young  stars $\rm
M^{(*)} _{ch}$  and thus on the  stellar IMF (Elmegreen  et al.  2008)
for the  ISM in CRDRs  are explored in  detail in a  forthcoming paper
(Papadopoulos et al. 2010).   It nevertheless worths pointing out that
the  large   boost  of  $\rm   T_k$(min)  in  CRDRs  in   an  (almost)
extinction-free manner across a  large range of densities in molecular
clouds cannot but have fundamental consequences on $\rm M^{(*)} _{ch}$
and the  emergent stellar IMF.   Indicatively for a $\rm  T_k$(min) in
UV-shielded  cores boosted  by a  factor of  10, the  Jeans  mass $\rm
M_{J}$$\propto$$\rm  T^{3/2}_k  n(H_2)^{-1/2}$ rises  by  a factor  of
$\sim $32!   (over an identical  density range), and  almost certainly
raises $\rm  M^{(*)} _{ch}$ and  the characteristic mass scale  of the
stellar  IMF.   Interestingly  similar   effects  can  occur  also  in
AGN-induced  X-ray  Dominated Regions  (XDRs;  Schleicher, Spaans,  \&
Klessen 2010) since X-rays just as CRs (and unlike far-UV photons) can
volumetrically heat large columns  of molecular gas while experiencing
very little  extinction.  The  effect of X-rays  on the Jeans  mass of
dense  cores and the  IMF has  been recently  proposed for  a powerful
distant QSO (Bradford  et al.  2009) and it  may represent a neglected
but important AGN feedback factor on its circumnuclear star formation.

\subsection{CRDRs:  observational tests}

The  molecular   line  diagnostics  of   starburst-induced  CRDRs  and
AGN-originating XDRs can be to a large degree degenerate when galaxies
host both power sources.  This has been noticed in earlier comparative
studies of  XDRs and regions  with higher $\rm U_{CR}$  values (though
only  up to 100$\times  $$\rm U_{CR,Gal}$)  found that  only carefully
chosen line ratios can distinguish between them (Meijerink, Spaans, \&
Israel 2006).  The still larger  CR energy densities expected in CRDRs
of ULIRGs will further compound these difficulties.

Provided that a powerful X-ray luminous AGN heating up the bulk of the
molecular gas  in its host galaxy  can be somehow  excluded (e.g.  via
hard X-ray observations),  any set of molecular lines  and ratios that
can strongly  constrain the temperature  of the dense  gas UV-shielded
phase ($\rm  n(H_2)$$>$10$^4$\,cm$^{-3}$) in ULIRGs  will be valuable.
Indeed, given that in  the hierachical structures of typical molecular
clouds the dense gas regions: a) lie well inside much larger ones that
strongly  attenuate far-UV light  and b)  cool strongly  via molecular
line emission because  of their high densities, then  any evidence for
high  temperatures for the  dense gas  phase would  be a  indicator of
strong  CR-heating.  In that  regard observations  of high-J  CO lines
such as J=6--5 {\it and}  its $^{13}$CO isotopologue have already been
proven    excellent    in     revealing    CR-heated    rather    than
UV/photoelectrically-heated   molecular   gas   in   galactic   nuclei
(Hailey-Dunsheath et  al.  2008).  Irrespective of  the particular set
of rotational lines used as  a ``thermometer'' of the dense gas, three
general requirements must be met: a) all lines must have high critical
densities  ($\rm   n_{crit}$$>$10$^{4}$\,cm$^{-3}$),  b)  have  widely
separated  $\rm  E_{J+1,J}/k_B$ factors,  and  c) the  J-corresponding
lines of  at least one rare  isotopologue must also  be observed (e.g.
$^{12}$CO {\it and} $^{13}$CO, or C$^{32}$S {\it and} C$^{34}$S, etc).
The first two ensure probing  of the dense gas phase while maintaining
good $\rm T_k$-sensitivity, and the last one is necessary for reducing
well-known  degeneracies when  modeling only  transitions of  the most
abundant  isotopologue which  often have  significant  optical depths.
Molecular lines with $\rm n_{crit}$$\ga $10$^5$\,cm$^{-3}$ (e.g.  HCN,
CS rotational transitions) are particularly valuable since, aside from
emanating from gas well within typical pre-stellar cores, they trace a
phase whose  kinematic state is  either dictated by  self-gravity (e.g
Goldsmith  2001), or has  fully dissipated  to thermal  motions.  This
constrains the  line formation  mechanism (and can  be used  to remove
degeneracies  of radiative  transfer  models e.g.   see  Greve et  al.
2009),  but most importantly  it reduces  the possibility  of residual
mechanical heating of  the dense gas (Loenen et  al.  2008) that could
mask as CR-heating (as both can heat gas volumetrically).

A  brief example of  such diagnostics  can be  provided using  a Large
Velocity  Gradient  (LVG)  code  (e.g.  Richardson  1985)  to  compute
relative line intensities for  dense gas with $\rm T_{k}$=(10--15)\,K,
and $\rm  n(H_2)$=10$^5$\,cm$^{-3}$, and $\rm  T_{k}$=(100--150)\,K at
the  same density.   In  both cases  a  gas velocity  gradient due  to
self-gravity  is assumed.   For the  cold  gas, the  CO and  $^{13}$CO
(J+1--J)/(3--2)    brightness    temperature    ratios    are:    $\rm
R_{(J+1,J)/32}\la  $0.7 for  J+1$\geq  $4 (J=1--0,  and  2--1 are  not
considered  because they  can  have substantial  contributions from  a
diffuse  non   self-gravitating  phase).    For  the  warm   gas  $\rm
R_{(J+1,J)/32}$$\sim $0.9--0.95 for  J+1=4--6, while the corresponding
ratios  for $  ^{13}$CO are  $\sim $1--1.3.   Similar  diagnostics but
using  multi-J   line  emission   from  rarer  molecules   (and  their
isotopologues)  with much  larger dipole  moments  ($\rm n_{crit}$$\ga
$10$^5$\,cm$^{-3}$)  such  as  HCN  and  H$^{13}$CN  can  even  better
constrain  the temperature of  dense gas  in galaxies.   However their
much fainter emission (e.g.  HCN  lines are $\sim $5--30 times fainter
than  those of  CO) allows  their use  only for  the  brightest nearby
starburst  nuclei  (e.g.   see  Jackson  et al.   1995  for  an  early
pioneering effort), and  only ALMA will enable such  diagnostics for a
large number of galaxies in the local and distant~Universe.

Strong thermo-chemical  effects induced  by large $\rm  U_{CR}$ values
and  their  ability to  volumetrically  warm  large  amounts of  dense
molecular gas can also provide valuable diagnostic of the temperatures
deep  inside  dense gas  cores.   For  example  global HNC/HCN  J=1--0
brightness temperature ratios of $\rm R^{(1-0)} _{HNC/HCN}$=0.5-1.0 in
galaxies  can be attributed  to an  ensemble of  PDRs but  ratios $\rm
R^{(1-0)} _{HNC/HCN}$$<$0.5  cannot, and imply  $\rm T_k$$\geq $100\,K
(Loenen et al.  2008).  Such low ratios are indeed found in ULIRGs but
are attributed  to turbulent  gas heating by  SNRs in  dense molecular
environments  (Loenen 2009) that  substantially warms  the gas  in the
absense  of photons  to  $\rm T_{k}$$\geq  $100\,K  necessary for  the
HNC+H$\rightarrow$HCN+H  reaction to  proceed efficiently  and convert
HNC  to HCN (e.g.   Schilke et  al.  1992).   Unlike the  Galaxy where
turbulent gas  heating has  subsided in the  dense subsonic  gas cores
that would emitt in these transitions this may not be so in the ISM of
ULIRGs, making CR and mechanical heating difficult to distinguish.  As
mentioned  earlier,  moving  the  CRDR molecular  line  diagnostic  to
tracers with  ever increasing critical densities  (and definitely with
$\rm  n_{crit}$$\ga  $10$^5$\,cm$^{-3}$)  may   be  the  only  way  of
``breaking''  this  degeneracy as  turbulent  heating  is expected  to
become progressively  weaker in higher  density cores (after  all star
formation is expected  to proceed in dense gas  cores whose turbulence
has fully dissipated, Larson 2005).  Finally, chemical signatures that
can uniquely  trace the  high $\rm U_{CR}$  values in CRDRs  using the
high  ionization fractions  expected  for their  dense  gas cores  are
particularly  valuable.  A  very sensitive  such probe  of  $\rm \zeta
_{CR}$ and  $\rm x(e)$  in dense UV-shielded  gas is provided  by $\rm
R_D$=$\rm   [DCO^+]/[HCO^+]$  versus   $\rm   R_H$=$\rm  [HCO^+]/[CO]$
abundance  ratio  diagrams where  the  high  CR  energy densities  and
resulting ionization  fractions in CRDRs would correspond  to very low
$\rm R_D$  and very high $\rm  R_H$ values (Caselli et  al.  1998).  A
dedicated  study of these  issues that  explores unique  diagnostic of
CRDRs is now in progress (Meijerink et al. 2010).

Most of  the aforementioned  tests will acquire  additional diagnostic
power once  ALMA will be able  to conduct them  at high, sub-arcsecond
resolution  in nearby  ULIRGs. It  will  then be  possible to  discern
whether  very  warm  dense  gas  with  high  ionization  fractions  is
localized  around a point-like  source, the  AGN (and  is thus  due to
XDRs), or  is well-distributed over  the star-forming area  of compact
systems and is thus starburst-related (see Schleicher et al.  2010 for
a  recent such  study).   High resolution  imaging  can also  directly
measure the  high brightness  temperatures expected for  all optically
thick   and  thermalized  molecular   lines  (where   $\rm  T^{(line)}
_{ex}$$\sim  $$\rm  T_k$)  in  CRDRs.  Radiative  transfer  models  of
emergent CO line emission from  a self-gravitating gas phase with $\rm
n(H_2)$=10$^5$\,cm$^{-3}$ and  $\rm T_k$=(100--150)\,K typically yield
$\rm T_b$(CO)$\sim $(80--140)\,K for  J=1--0 up to J=7--6, and imaging
at a linear resolution of $\rm \Delta L$$\la $100\,pc (the diameter of
gaseous disks  in ULIRGs) would  be adequate to directly  measure them
(see  Sakamoto   et  al.   2008;   and  Matshushita  2009   for  early
examples).  For z$\la  $0.05 (which  would include  a large  number of
ULIRGs) such  linear resolutions correspond to  angular resolutions of
$\rm  \theta  _{b}$$\sim $0.1$''$  (for  a  flat $\Lambda  $-dominated
cosmology with $\rm H_0=71\,$km\,s$^{-1}$\,Mpc$^{-1}$ and $\Omega_{\rm
m}=0.27$) which would be certainly possible with ALMA.

\newpage

\section{Conclusions}

The main conclusions of this work are three, namely

\begin{itemize}

\item In the high star-formation density environments of ULIRGs cosmic
ray energy  densities $\rm U_{CR}$  will be enhanced by  a tremendeous
factor  of $\sim $few$\times$(10$^3$--10$^4$)$\rm  U_{CR,Gal}$.  These
will  permeate the  large molecular  gas reservoirs  of  such systems,
likely turning  them into CR-dominated  regions (CRDRs) where  CRs not
far-UV photons regulate  the thermal and ionization state  of the bulk
of their typicaly very dense molecular gas.

\item Irrespective whether the  tremendeous CR energy densities in the
compact starburst  regions of ULIRGs provide the  dominant heating for
most of their  molecular gas or not, they  will dramatically raise the
gas temperature and ionization  fraction {\it minimum} values possible
in their  ISM, which are  typically attained in UV-shielded  dense gas
cores where the star-formation initial conditions are set.

\item The new and very different initial conditions for star formation
        in  CRDRs   {\it  are  an  imperative   for  all  high-density
        star-formation  events,} and will  almost certainly raise  the
        characteristic mass  of young stars (and thus  the stellar IMF
        mass scale) during such events throughout the Universe.

\end{itemize}

Sensitive  observations  of key  molecular  lines  with high  critical
densities  ($\rm  n_{crit}$$\ga $10$^5$\,cm$^{-3}$)  as  well as  high
resolution  mm/submm  interferometric imaging  will  be  key tools  in
uncovering the high temperatures of the dense UV-shielded gas in CRDRs
expected  in compact  extreme starbursts.   Tracers of  the  very high
ionization  fractions expected  for  their dense  gas  can provide  an
independent assesment of their  presense.  All these tests will become
possible with ALMA for large numbers of ULIRGs in the local Universe.

\acknowledgements  It is  a pleasure  to acknowledge  many informative
discussions with  Drs Wing-Fai Thi and  Francesco Miniati. Discussions
with Dr Marco Spaans  regarding XDR/CDR potential diagnostics are also
greatfuly acknowledged.  Finally I would like to thank the referee for
his/her extensive  set of comments that helped  helped clarify several
issues and significantly  widen the scope of this work.

\newpage

\begin{figure}
\epsscale{1.1}
\plotone{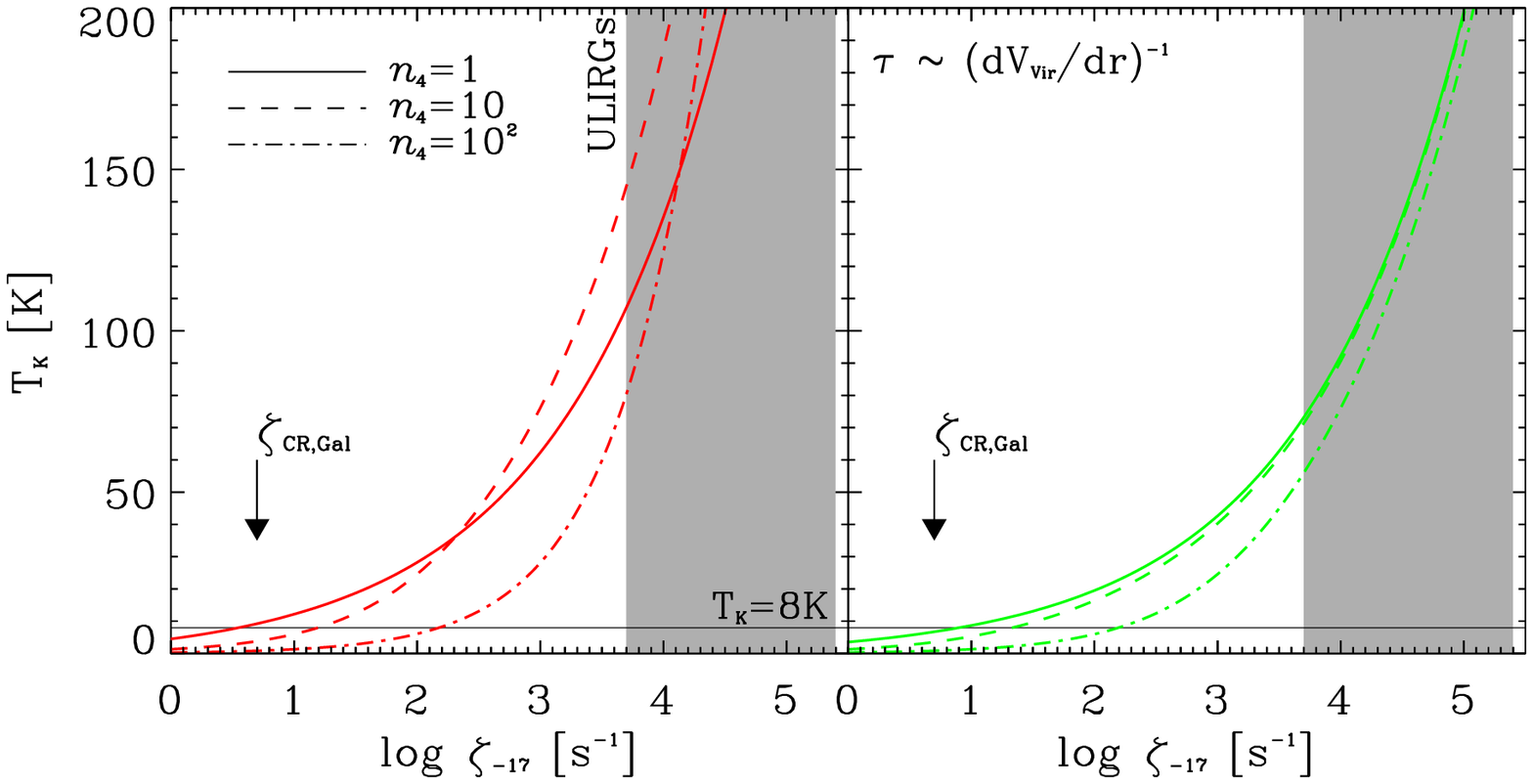}
\caption{The temperatures  of UV-shielded gas  in CR-dominated Regions
(CRDRs)  of compact  extreme starbursts  in ULIRGs  (marked  by shaded
area)        for        cores        with        densities        $\rm
n(H_2)$=(10$^4$,10$^5$,10$^6$)\,cm$^{-3}$  and  near  thermal  motions
(left), or dictated by gas self-gravity (right) that yield larger than
thermal linewidths  and stronger line cooling (see  section 2.1).  The
horizontal line marks the typical temperature of this gas phase in the
Galaxy. The arrow marks the adopted Galactic CR ionization rate of $\rm
\zeta _{CR,Gal}$=$5\times 10^{-17}$\,s$^{-1}$. }
\end{figure}

\newpage

\begin{figure}
\epsscale{1.0}
\plotone{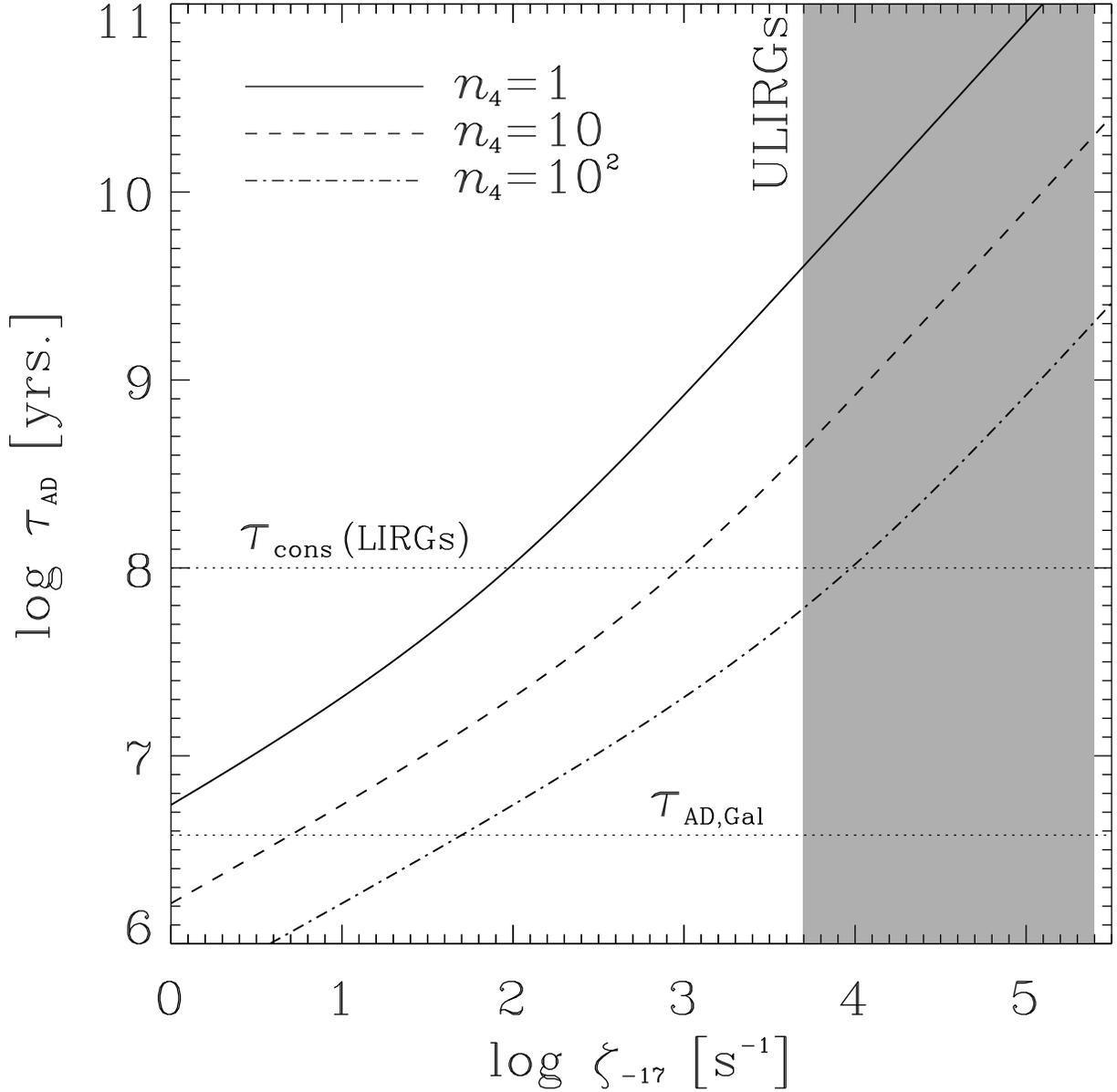}
\caption{The  ambipolar  diffusion  timescale  of UV-shielded  gas  in
CR-dominated Regions  (CRDRs) of compact extreme  starbursts in ULIRGs
(marked   by   shaded   area)    for   cores   with   densities   $\rm
n(H_2)$=(10$^4$,10$^5$,10$^6$)\,cm$^{-3}$ estimated  from Equation 17.
The  lower horizontal  line marked  by $\rm  \tau _{AD,  Gal}$  is the
ambipolar   diffusion  timescale  for   UV-shielded  cores   with  $\rm
n(H_2)$=$10^5$\,cm$^{-3}$    in   the    Galaxy    for   $\rm    \zeta
_{CR,Gal}$=$5\times 10^{-17}$\,s$^{-1}$, and the higher line marks the
gas consumption timescale  of the entire molecular gas  reservoir of a
typical LIRG ($\rm L_{IR}$$\sim $10$^{11}$\,L$_{\odot}$). For the more
vigorously star-forming  ULIRGs and  their HCN-bright dense  gas phase
this  gas consumption  timescale can  be  an order  of magnitude  less
(section 3).}
\end{figure}

\end{document}